\documentclass[11pt]{article}
\usepackage{Blois,epsfig}

\def\be{\begin{equation}}
\def\ee{\end{equation}}
\def\bea{\begin{eqnarray}}
\def\eea{\end{eqnarray}}

\begin{document}
\vspace*{2cm}
\begin{center}
\Large{\textbf{XIth International Conference on\\ Elastic and Diffractive Scattering\\ Ch\^{a}teau de Blois, France, May 15 - 20, 2005}}
\end{center}

\vspace*{2cm}
\title{THE TOTAL CROSS SECTION AT THE LHC}

\author{ P V LANDSHOFF}

\address{Department of Applied Mathematics and Theoretical Physics,\\
University of Cambridge, Wilberforce Road,\\
Cambridge CB3 0WA, England}

\maketitle\abstracts{
The hard pomeron first came to light in deep inelastic lepton scattering,
but evidence that it contributes also to soft hadronic collisions is
reinforced by the fact that it seems to obey a factorisation similar
to that of other Regge exchanges. Including a hard-pomeron term 
in a fit to data for $\sigma^{pp}$ and $\sigma^{p\bar p}$ leads to a
large total cross section at the LHC. An exact prediction is not possible
because of the uncertaintities arising from screening corrections; the best
estimate is $125\pm25$ mb.}

\section{The LHC total cross section}\label{subsec:prod}

Most predictions of the total cross section at the LHC are at the level of
100~mb or a little larger\cite{cudell}. I am going to argue that it may in fact be
significantly larger, as much as  125 or even 150~mb. 

Nearly 15 years ago, Donnachie and I fitted\cite{DL1} all hadronic 
total cross section data with the simple Regge form\cite{book}
\be
\sigma=X_1s^{\epsilon_1}+Ys^{\epsilon_2}
\label{soft}
\ee
Here $\epsilon_1\approx 0.08$ and $\epsilon_2\approx -0.5$. The second term
represents $\rho,\omega,f_2,a_2$ exchange and the first term, whose dynamical
origin is still poorly understood, is said to correspond to ``soft-pomeron''
exchange. The form (\ref{soft}) correctly predicted the $\gamma p$ total
cross section that was measured shortly afterwards at HERA. The extension
to $\gamma^* p$ collisions would be that, at small $x$,
\be
F_2(x,Q^2)\sim A_1(Q^2)\,x^{-\epsilon_1} +A_2(Q^2)\,x^{-\epsilon_2}
\label{deepin}
\ee
but this did not fit the subsequent HERA data. However\cite{DL2}, 
an excellent fit to the data is obtained by adding to (\ref{deepin})
a term $A_0(Q^2)\,x^{-\epsilon_0}$ with $\epsilon_0\approx 0.4$. We call this
term ``hard-pomeron'' exchange. 

The obvious question is whether a hard-pomeron exchange term
$X_0s^{\epsilon_0}$ should be added to (\ref{soft}). Up to CERN SPS
collider energies, the form (\ref{soft}) fits the data very well, but
there are two conflicting measurements\cite{tev} of $\sigma(p\bar p)$ at the
Tevatron. If the larger (CDF) result is correct, it is a clear indication
of a new effect, which can be well represented by the hard-pomeron term.

A simultaneous fit\cite{DL3}
to data for $\sigma(pp),\sigma(p\bar p),\sigma(\gamma p)$
and $F_2$, with hard-pomeron exchange included for each
process, is shown in figures 1 and 2a. Belief in the fit is strengthened by the
fact that Regge factorisation allows one to extend it, with no further
free parameters, to data for $\sigma(\gamma \gamma), \sigma(\gamma^* \gamma)$
and $\sigma(\gamma^* \gamma^*)$. I discuss this in the next section.
\begin{figure}[t]
\begin{center}
{\epsfxsize=0.45\hsize\epsfbox[125 495 470 770]{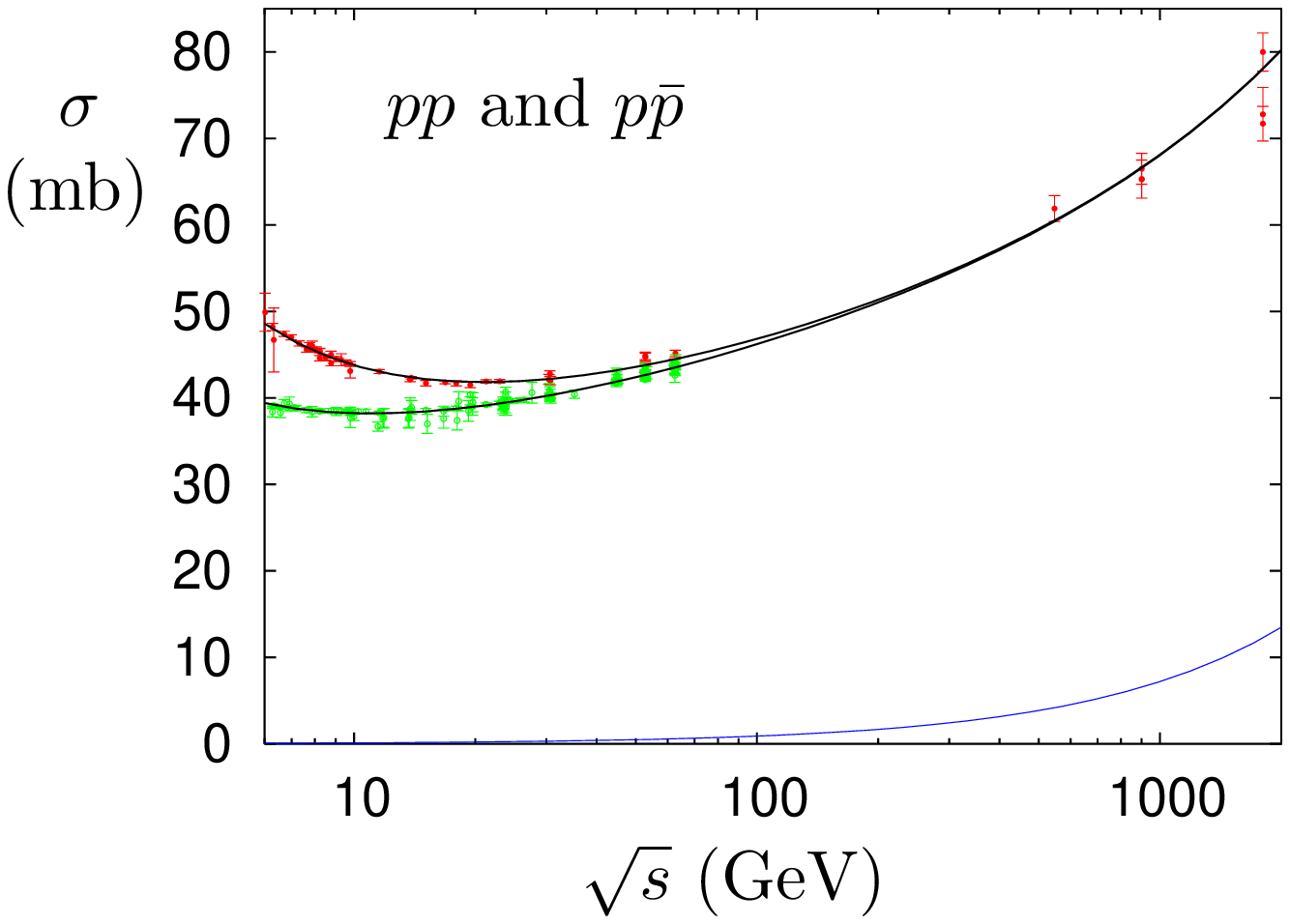}
\hfill
\epsfxsize=0.45\hsize\epsfbox[125 495 470 770]{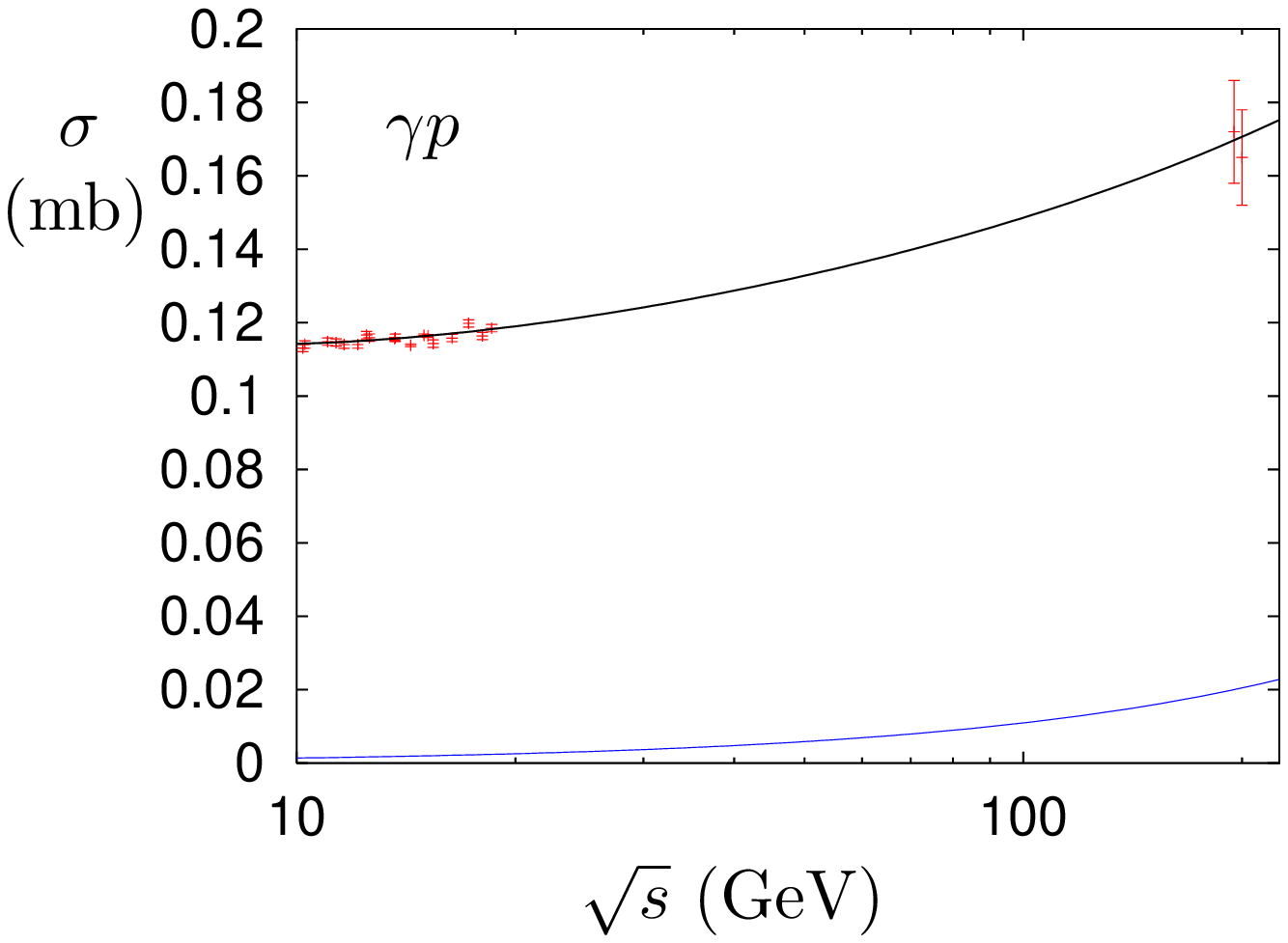}}
\end{center}
\caption{Fits to $\sigma(pp),\sigma(p\bar p),\sigma(\gamma p)$; in each case the
lower line is the hard-pomeron contribution}
\label{SCATTERING}
\vskip 8truemm
\begin{center}
{\epsfysize=0.5\vsize\epsfbox[65 295 377 755]{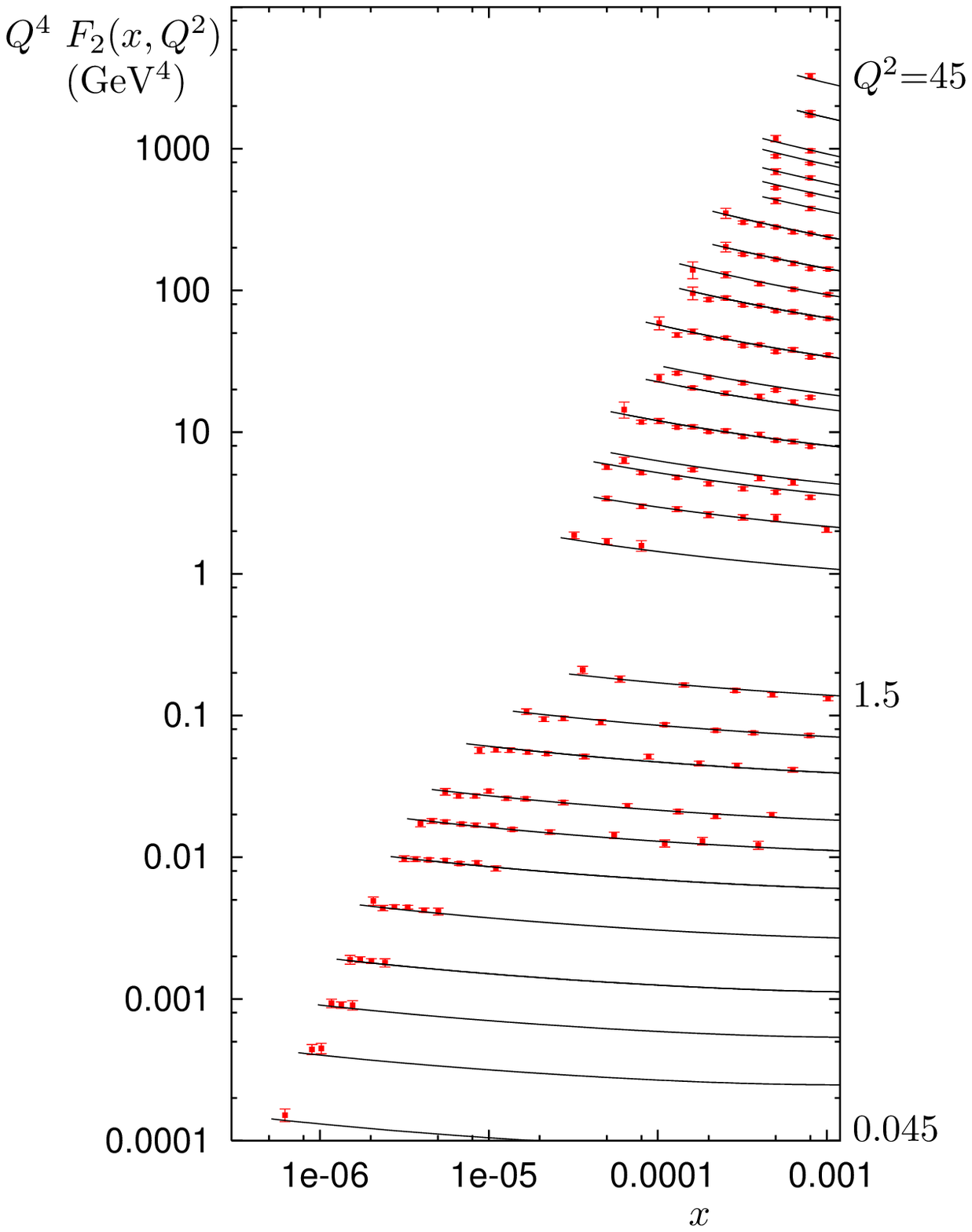}\hfill
\epsfysize=0.6\hsize\epsfbox[65 295 377 755]{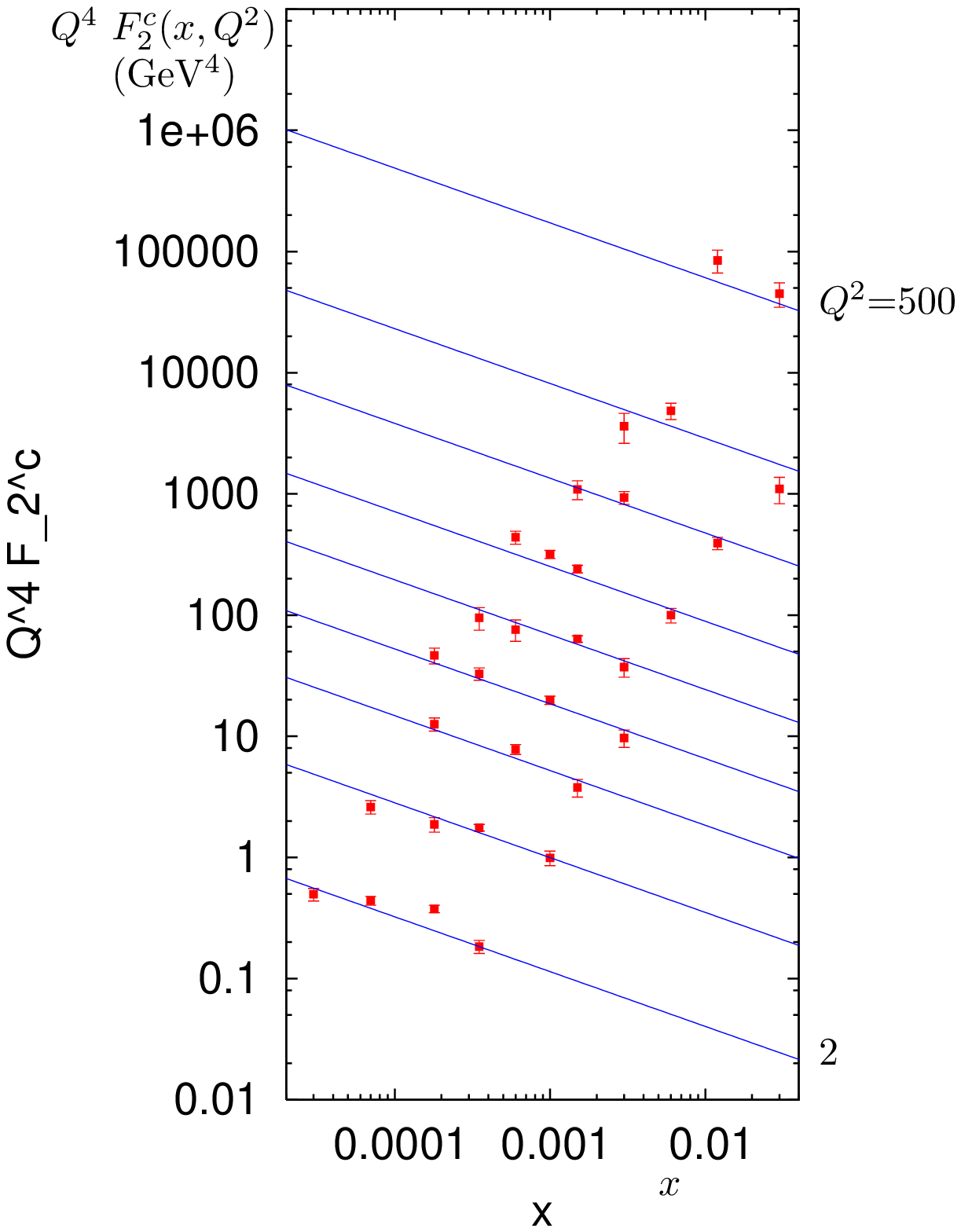}}

{\hfill (a)\hfill (b)\hfill\hfill}
\end{center}
\caption{(a) Fit to the proton structure function; (b) 
its charm component with lines corresponding to hard-pomeron
exchange.}
\end{figure}

\begin{figure}[t]
\begin{center}
{\epsfxsize=\hsize\epsfbox[80 560 530 770]{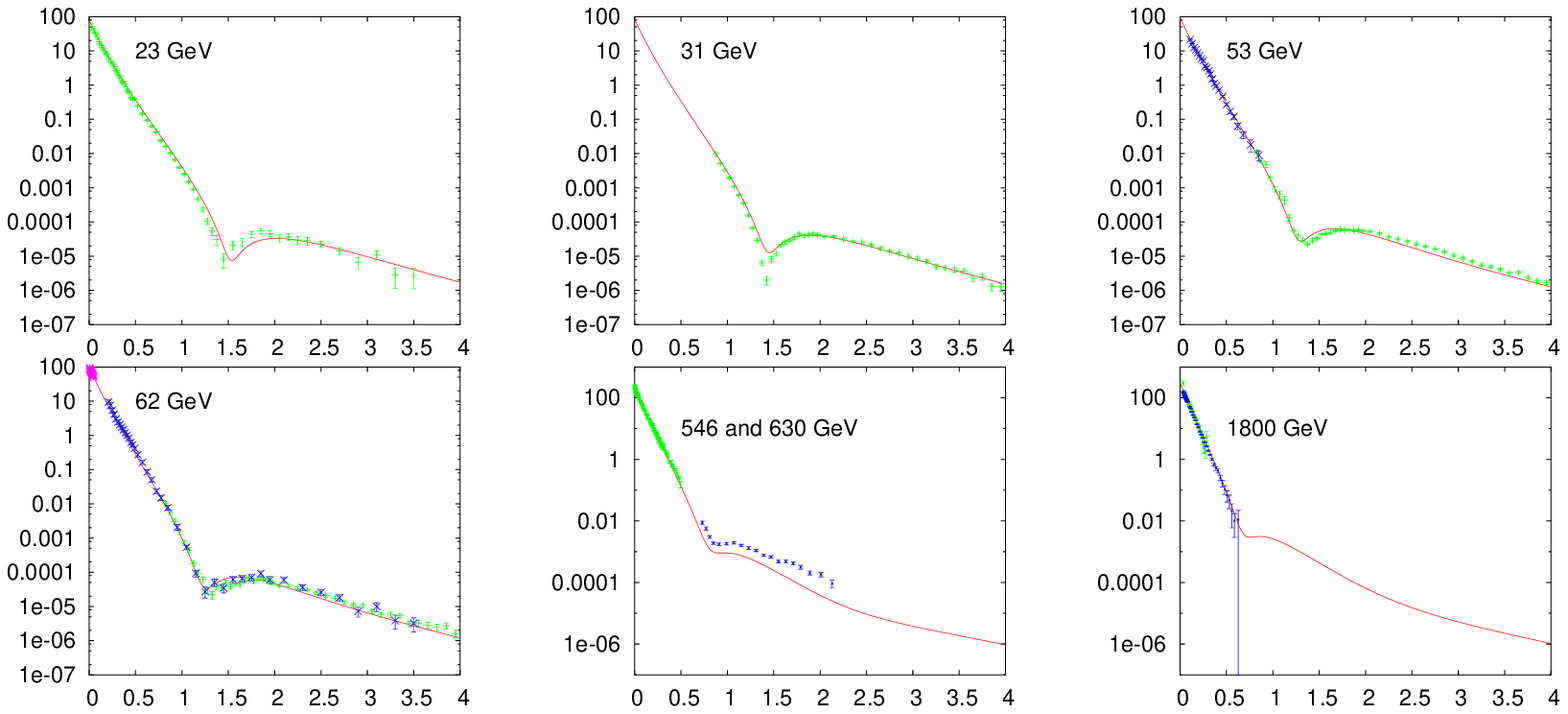}}
\end{center}
\caption{Fits to $pp$ and $p\bar p$ elastic scatterin}
\vskip 8truemm
\begin{center}
{\epsfxsize=0.44\hsize\epsfbox[110 500 485 775]{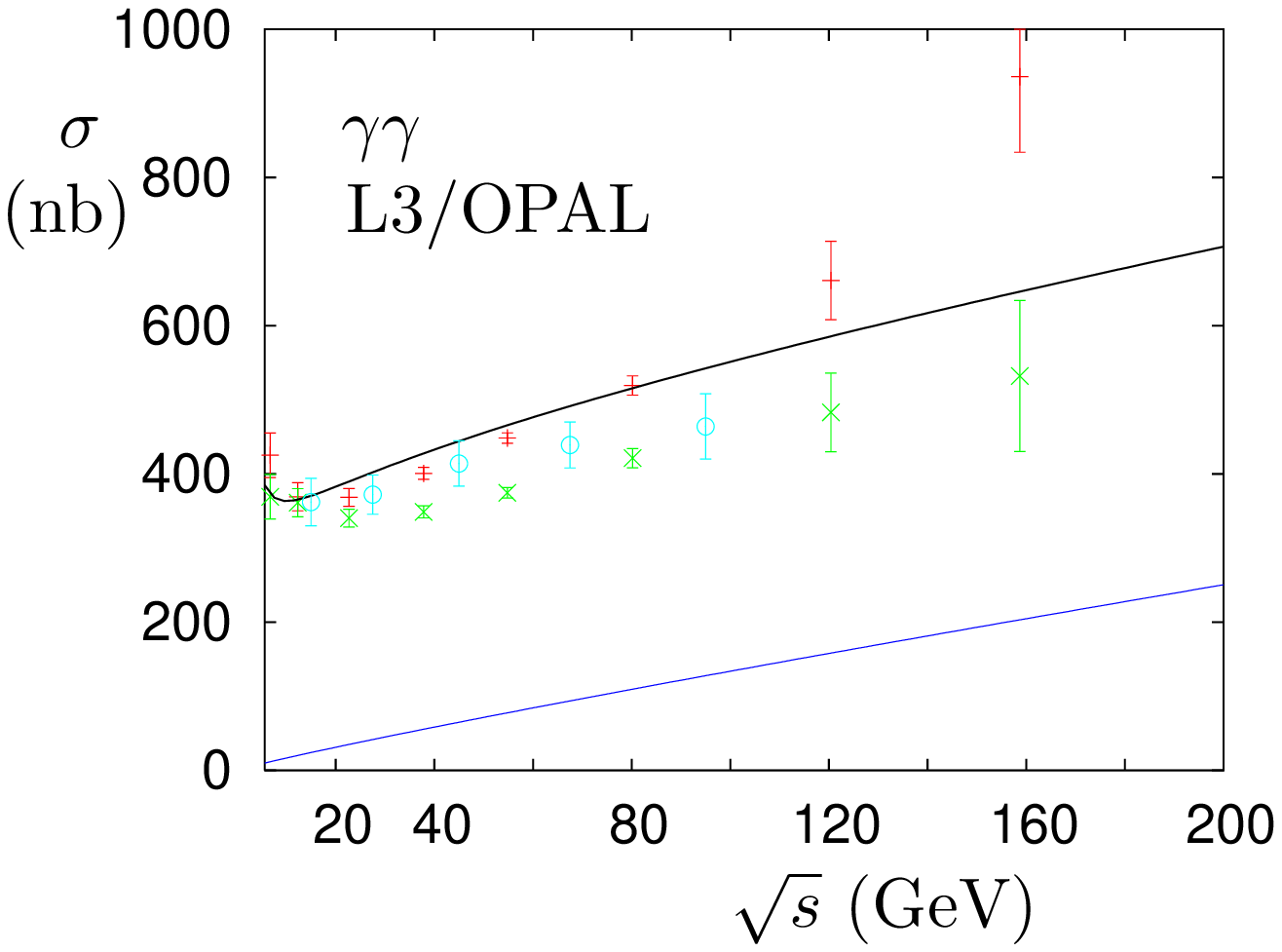}\hfill
\epsfxsize=0.4\hsize\epsfbox[130 500 470 755]{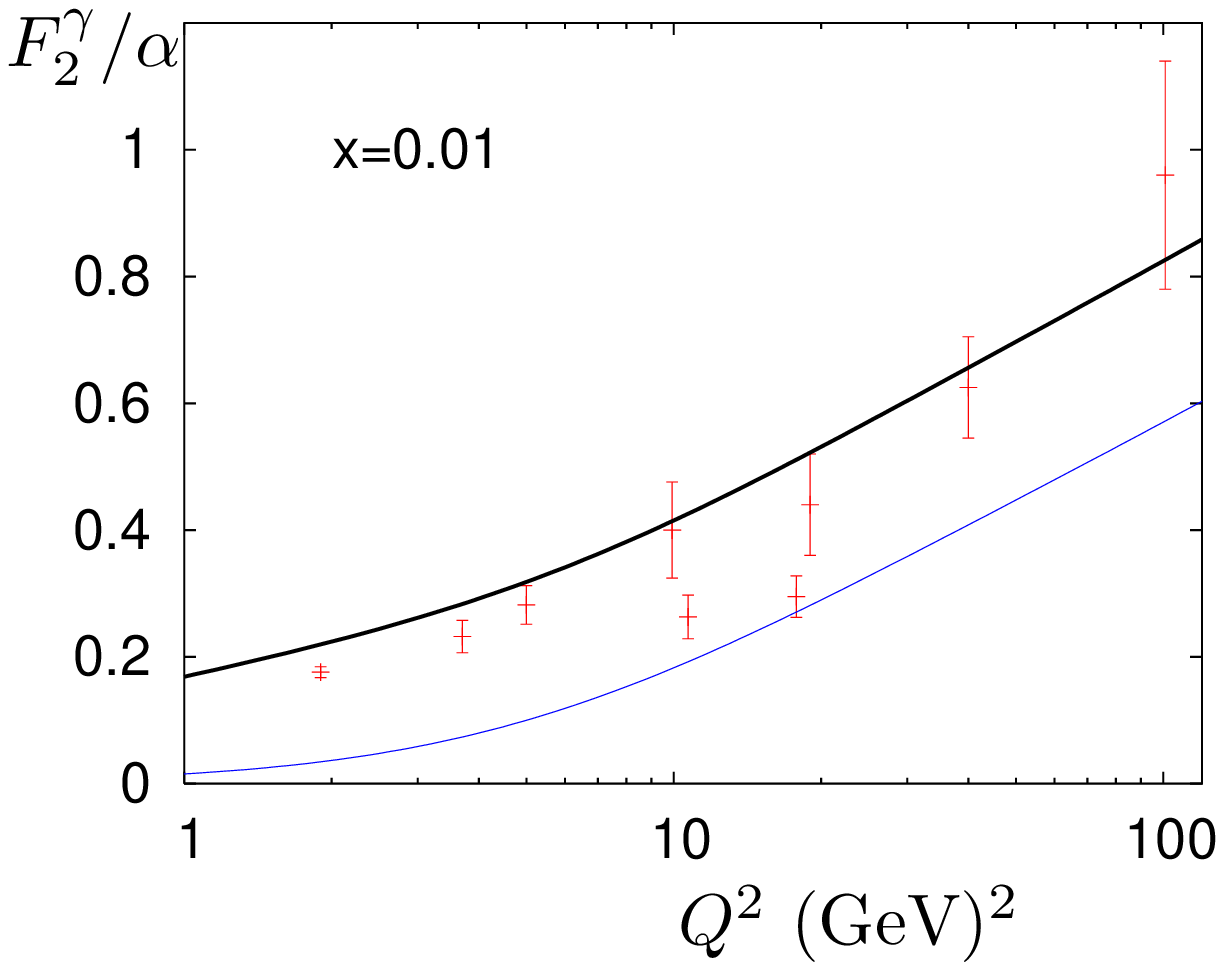}}
\vskip 4truemm
{\epsfxsize=0.5\hsize\epsfbox[50 500 470 770]{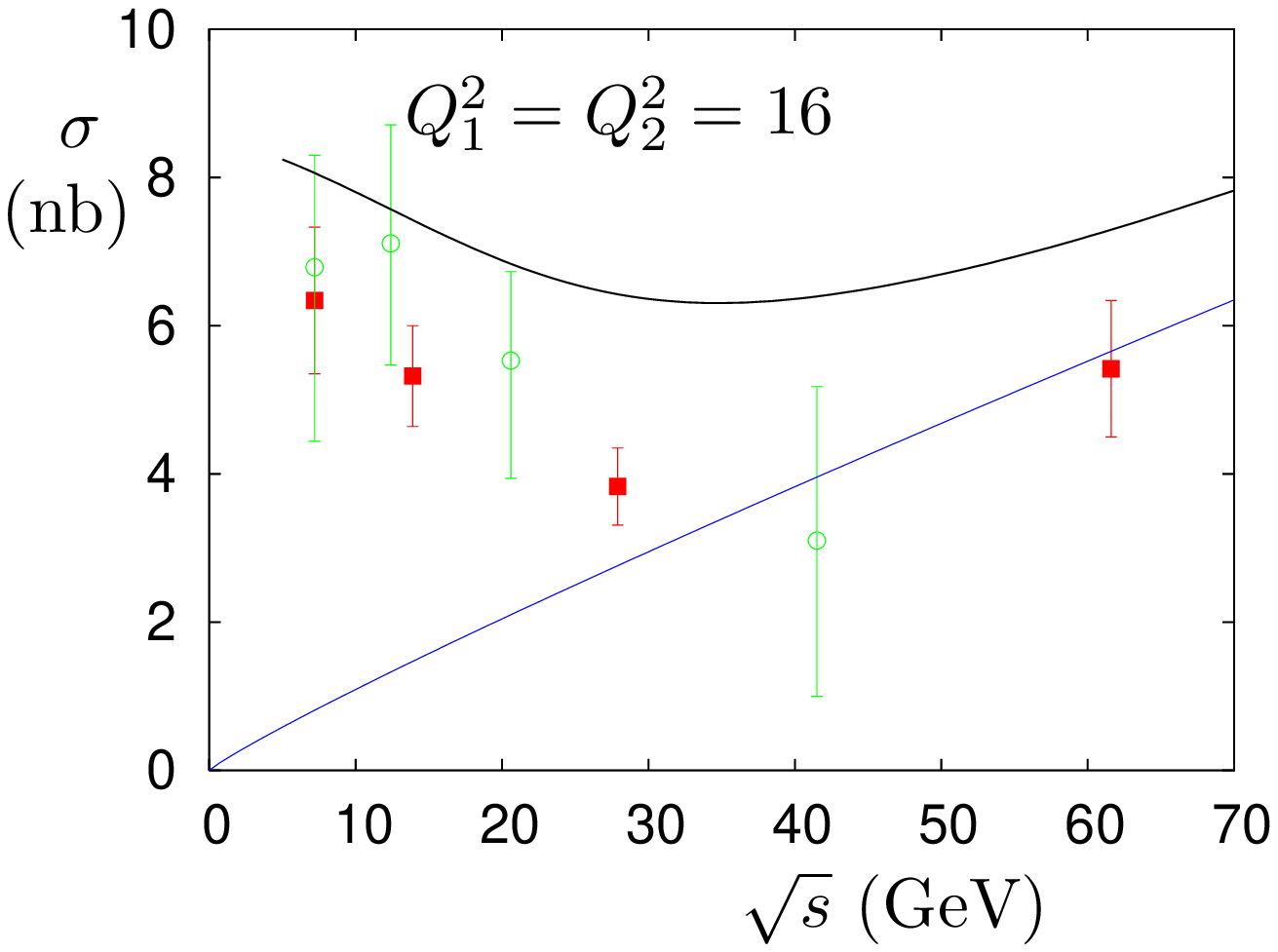}}
\end{center}
\caption{Tests of Regge factorisation. In each case the lower line is
the hard-pomeron contribution.}
\end{figure}

\begin{figure}[t]
\begin{center}
{\epsfxsize=0.44\hsize\epsfbox[110 500 485 775]{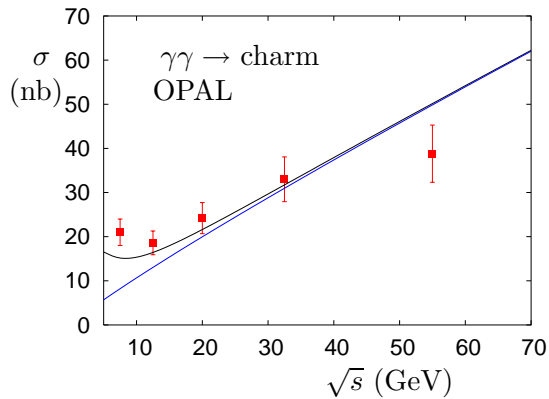}}
\end{center}
\caption{Prediction from Regge factorisation for 
charm production in $\gamma\gamma$ collisions. The lower curve is the contribution from hard-pomeron exchange; the upper curve includes also the box graph.}
\end{figure}

An extrapolation to the fit to LHC energies gives 150~mb or more. 
However, this is without shadowing corrections, that is the simultaneous
exchange of more than one pomeron. But nobody knows how to calculate this.
The best one can do is to follow a procedure that Donnachie and I
introduced many years ago\cite{DL4} and fix the magnitude of the double 
exchange from elastic-scattering data. If one calculates the 
impact-parameter-space
amplitude $a(s,b)$ corresponding to single exchange, then
$-\lambda [a(s,b)]^2$, with $\lambda$ real, has\cite{book}
the correct analytic properties to represent double exchange. 
Fix $\lambda$ by requiring that the dip discovered
in CERN ISR $pp$ elastic scattering occurs at the correct value of $t$
(which needs $\lambda\approx 0.4$). There is no reason to believe that this
gives the correct amplitude, but it is the best one can do.

Including this extra term in the fits to $\sigma(pp),\sigma(p\bar p)$ gives
a slightly-improved fit. For $pp$ and $p\bar p$ elastic scattering we must also
include triple-gluon exchange\cite{DL4}. The resulting curves are shown in 
figure~3. They are reasonably good, but only reasonably good. Because of
this, there is a large error in the predicted value of $\sigma(pp)$ at
the LHC; hence my prediction 125$\pm 25$~mb.

\section{Regge factorisation}\label{subsec:regge}

If the three exchanges obey Regge factorisation, then the contribution
of each of them to the cross sections satisfies
\be
\sigma(\gamma\gamma)={\sigma(\gamma p)\sigma(\gamma p)/\sigma(pp)}
\hbox{ for all }Q_1^2,Q_2^2
\label{factor}
\ee
This has been quite well tested for soft-pomeron exchange, but for
hard-pomeron exchange the position is less clear\cite{DL3}.
To calculate $\sigma(\gamma\gamma)$ one must also include the box graph.

Fixing the parameters from the fits of figures 1 and 2a gives, when we
use (\ref{factor}) and assume that the screening corrections can be
neglected, zero-parameter predictions for $\sigma(\gamma\gamma)$.
See figure~4.
Evidently the data are not good enough to enable any clear conclusion
to be reached, though the factorisation hypothesis does appear to be
consistent with the data.

Some further evidence comes from charm production. The part of $F_2$
corresponding to events in which a charmed particle is produced is
dominated by hard-pomeron exchange alone at small $x$; for some reason
the hard pomeron does not couple. This is seen in figure 2b. If we
apply Regge factorisation we arrive at the prediction
for the charm production in $\gamma\gamma$ collisions shown in figure 5. 
Again the data are
not good enough to reach a firm conclusion, but the factorisation hypothesis 
does seem to work quite well.

\end{document}